\documentclass[twoside,magazine]{IEEEtran}
\usepackage{makecell}
\usepackage{hyperref}
\usepackage{array}
\usepackage{graphicx,amssymb,amsmath}
\usepackage{multicol}
\usepackage[noadjust]{cite}
\usepackage{setspace}
\usepackage{subfig}
\usepackage{graphicx}
\usepackage{float}
\usepackage {url}
\usepackage{stfloats}
\usepackage{amsthm,pifont}
\usepackage{flushend}
\usepackage{cases,subeqnarray}
\usepackage{bm,multirow,bigstrut}
\usepackage{amsmath, amsthm, amssymb}
\usepackage{textcomp}
\usepackage{latexsym,bm}
\usepackage{booktabs}
\usepackage{xcolor}
\usepackage{mathtools}
\usepackage{dsfont}
\usepackage{extarrows}
\usepackage{epsfig}
\usepackage{epsfig}
\usepackage{epstopdf}
\usepackage[noend]{algpseudocode}
\usepackage{algorithmicx,algorithm}
\usepackage{xspace}
\usepackage{listings}
\usepackage{xcolor}
\definecolor{commentcolor}{RGB}{85,139,78}
\definecolor{stringcolor}{RGB}{206,145,108}
\definecolor{keywordcolor}{RGB}{0,0,128}
\definecolor{backcolor}{RGB}{220,220,220}
\theoremstyle{plain}

\theoremstyle{plain}

\usepackage{amsmath}

\IEEEoverridecommandlockouts

\begin{document}
\title{Wireless Sensing Data Collection and Processing for Metaverse Avatar Construction}
\author{Jiacheng Wang, Hongyang Du, Xiaolong Yang,~\IEEEmembership{Member,~IEEE}\\ Dusit~Niyato,~\IEEEmembership{Fellow,~IEEE}, Jiawen~Kang, and Shiwen Mao,~\IEEEmembership{Fellow,~IEEE},

}

\maketitle
\vspace{-1cm}
\begin{abstract}
Recent advances in emerging technologies such as artificial intelligence and extended reality have pushed the Metaverse, a virtual, shared space, into reality. In Metaverse, users can customize virtual avatars to experience a different life. While impressive, avatar construction requires a lot of data that manifest users in the physical world from various perspectives, and wireless sensing data is one of them. For example, machine learning (ML) and signal processing can help extract information about user behavior from sensing data, thereby facilitating avatar behavior construction in the Metaverse. This article presents a wireless sensing dataset to support the emerging research on Metaverse avatar construction. Rigorously, the existing data collection platforms and datasets are analyzed first. On this basis, we introduce the platform used in this paper, as well as the data collection method and scenario. We observe that the collected sensing data, i.e., channel state information (CSI), suffers from a phase shift problem, which negatively affects the extraction of user information such as behavior and heartbeat and further deteriorates the avatar construction. Therefore, we propose to detect and correct this phase shift by a sliding window and phase compensation, respectively, and then validate the proposed scheme with the collected data. Finally, several research directions related to the avatar construction are given from the perspective of datasets.

\end{abstract}
\IEEEpeerreviewmaketitle
\section{Introduction}


With the support of several technologies, e.g., computer science, artificial intelligence, wireless communications, as well as human-computer interaction, the Metaverse is constructed as a virtual world that would significantly change people's lives in the near future~\cite{wang2022survey}. Instead of browsing information on a screen today, in the Metaverse, various immersive services will be provided, which stimulate the user's senses and make the transmission of information more efficient. Furthermore, many experiments that are impossible or challenging to perform in the real world, can now be carried out in the Metaverse. For instance, with the assistance of devices such as augmented reality (AR), users can travel the world at home with friends who are located thousands of miles away. Therefore, it is believed that Metaverse embodies people's longing for the next generation of the Internet or even beyond.

The most essential feature of Metaverse services is ``Human-Centric''~\cite{du2022exploring}. From one perspective, Metaverse services require a deeper level of user involvement. With the support of virtual reality (VR), AR, and tactile Internet, Metaverse hardware devices can not only engage all of the user's senses to offer an immersive experience~\cite{wang2022survey}, but also revolutionize how people interact with one another, and even with objects. In addition, quality-of-experience (QoE) of users is an important performance metric in Metaverse service design~\cite{du2022exploring}. Different from the conventional online services, Metaverse designers not only need to pay attention to objective indicators such as latency and bit error rate, but also has to consider holistically detailed subjective factors of users in the physical world, such as location, status, behavior, and even psychological factors and attention. Therefore, any factors that may cause discomfort to the user needs to be circumvented. For instance, in VR-based Metaverse services, for tracking the user's behavior for virtual avatar synchronization in the Metaverse, non-intrusive wireless sensing will be a better solution than the wearable sensors.


Therefore, as one of the driving engines, wireless sensing should complete a variety of sensing tasks to satisfy the requirements of the human-centric Metaverse processes. Large scale services, such as virtual exhibitions, involve tracking of users' movements in the real world, and requires sensing tasks include target detection, activity recognition, and localization. At a small scale, Metaverse service, such as virtual meeting, needs to perform user emotion recognition and gesture interaction. Thus, sensing tasks such as behavior recognition, breathing, and heartbeat detection are vital components. Fortunately, in recent years, the practicality and accuracy of wireless sensing have increased with the continuous advancement of machine learning (ML)-based algorithms. Together with well developed signal processing algorithms, it is becoming highly feasible to support various Metaverse services with wireless sensing technology.

To fully achieve the Metaverse vision, one of the remaining obstacles is that the training of artificial intelligence (AI) models and the testing of ML algorithms require large amounts of sensing data, such as channel state information (CSI) data, which includes amplitude and phase information. Both can be used to extract information related to users' behavior in the physical world. Yet for sensing tasks with different levels of granularity, a part of the CSI information could be appropriately selected for processing. For instance, the tasks that require a high degree of accuracy, such as heartbeat analysis, the phase information is typically more important than the amplitude information. To provide a stronger data support for researches, various platforms~\cite{halperin2011tool,xie2015precise, gringoli2019free} and the corresponding datasets~\cite{ma2018signfi,zheng2019zero,qian2017widar} have been proposed. However, the majority of the datasets are gathered via 802.11n protocol based devices, which has a constrained signal bandwidth and limited number of antennas, making it challenging to meet the diverse research needs. Motivated by this, we propose a novel method in this paper for obtaining stable CSI data from an 802.11ac-based platform, as well as an open-source dataset, for avatar construction related researches. The contributions are summarized as follows:

\begin{figure*}[t]
	\centering
	\includegraphics[width=0.96\textwidth]{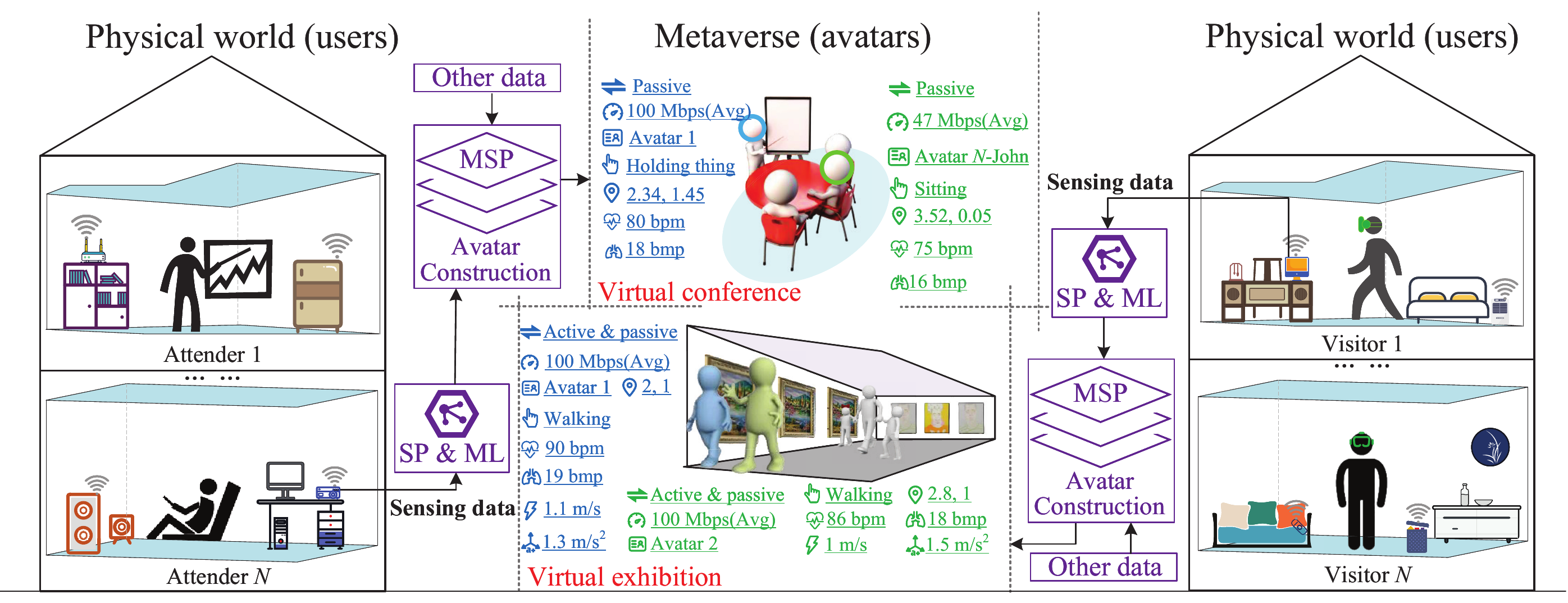}%
	\caption{The general framework of using wireless sensing data to support the construction of avatars in virtual conference and virtual exhibition applications.}
	\label{model}
\end{figure*}

\begin{itemize}
\item We present an application of wireless sensing technology in the construction of Metaverse avatars. To the best of our knowledge, this is the first work to propose the idea of applying wireless sensing to Metaverse services.

\item We present a general framework for the construction of Metaverse avatars by using the wireless sensing data through examples.

\item We conducted real-world experiments and made our database publicly available~\cite{xx2c-zg31-22}
Our data, which covers a larger bandwidth and involves more antennas, can be used to validate various wireless sensing algorithms related to avatar construction.

\end{itemize}

\section{Wireless sensing assisted avatar construction}
Construction of the Metaverse avatar requires a lot of data that depicts the users in the physical world at different scales and dimensions. The ubiquitous wireless signals and related sensing technologies provide strong support for it. In Fig.~\ref{model}, we use virtual conference and virtual exhibition as examples to illustrate the overall framework of wireless sensing data supports avatar construction in the Metaverse. Conceretly, the framework contains three core parts. First, in the physical world, the smart devices sense each user by transmitting and receiving wireless signals. Then, the signal processing (SP) and machine learning (ML) modules process the collected sensory data to extract various types of information about the user. Finally, this information is fed to the Metaverse service provider (MSP), which combines the sensing data with other types of information, such as that extracted from the image and sound sensors, to complete the construction and rendering of the avatar.
\subsection{Virtual conference}
The general procedure for using wireless sensing data to support avatar construction is described above. In fact, different Metaverse services have different sensing preferences. For example, in a virtual meeting, large-scale activities are less likely to take place and each user's range of movement is restricted, while the body language is relatively rich. Under such a case, the construction of virtual avatar might require more fine-grained user information, such as user behaviors and gestures. Therefore, SP and ML modules need to focus more on gathering user-specific small-scale features. In addition, the user's vital signs, such as breath, heartbeat, and other subtle physiological characteristics are also essential. These indicators can be used to analyze the user's mentality, emotion, etc., so as to build a more vivid avatar in the Metaverse.
\subsection{Virtual exhibition}
Unlike in virtual conferences, users tend to have more large-scale activities in virtual exhibitions, since they need to walk around. In this case, the information about the user's activity, location, movement direction, speed, etc., is more important for the virtual avatar construction in the Metaverse. For instance, the exhibited works seen at various locations throughout the exhibition are different, making the avatar's location is somewhat more significant than the gesture. Therefore, the SP and ML module should concentrate on the extraction of large-scale information. In addition, similar to the virtual meeting case, we believe that the user's breathing and heartbeat are also critical, as they can be used to evaluate the user's physical state, based on which the MSP can make personalized notifications for different avatars. 

From the above analysis, it is not difficult to see  that, first, wireless signals are ubiquitous in our daily lives, indicating the data sources are sufficient. Second, wireless signals can be used to complete various sensing tasks at different scales, demonstrating the data is valuable. Therefore, from both quantitatively and qualitatively, the sensing data is indispensable for avatar construction. 

\section{Testbeds and datasets}
In this section, we review several existing open-source test platforms and available datasets.
\subsection{The existing open-source test platforms}
There are several experimental platforms that can be used to collect CSI data. The authors in~\cite{halperin2011tool} provide the 802.11n CSI Tool. By using this tool, CSI can be collected as 30 complex numbers, each of which holds 8-bit signed real and imaginary parts. Another tool is called Atheros CSI Tool, which is build upon the open-source kernel driver of ath9k~\cite{xie2015precise}. However, neither of these supports the collection of CSI data with a bandwidth of 80 MHz. Fortunately, the Nexmon CSI extractor is introduced in~\cite{gringoli2019free}, enabling per-frame CSI extraction for up to four spatial streams using up to four receive chains on modern Broadcom and Cypress Wi-Fi chips with up to 80 MHz bandwidth in both the 2.4 and 5 GHz bands. The tool can be installed on devices ranging from the low-cost Raspberry Pi to mobile platforms such as Nexus smartphones, and an off-the-shelf Wi-Fi access points (AP). Later in~\cite{gringoli2022ax}, Nexmon is extended to the IEEE 802.11AX protocol based plafrom, realizing the extraction of CSI with the bandwidth of 160 Mhz. Additionally, the software-defined radio (SDR) platform can be utilized to obtain CSI, but the hardware cost is relatively high. 

\begin{figure*}[t]
	\centering
	\includegraphics[height=14cm]{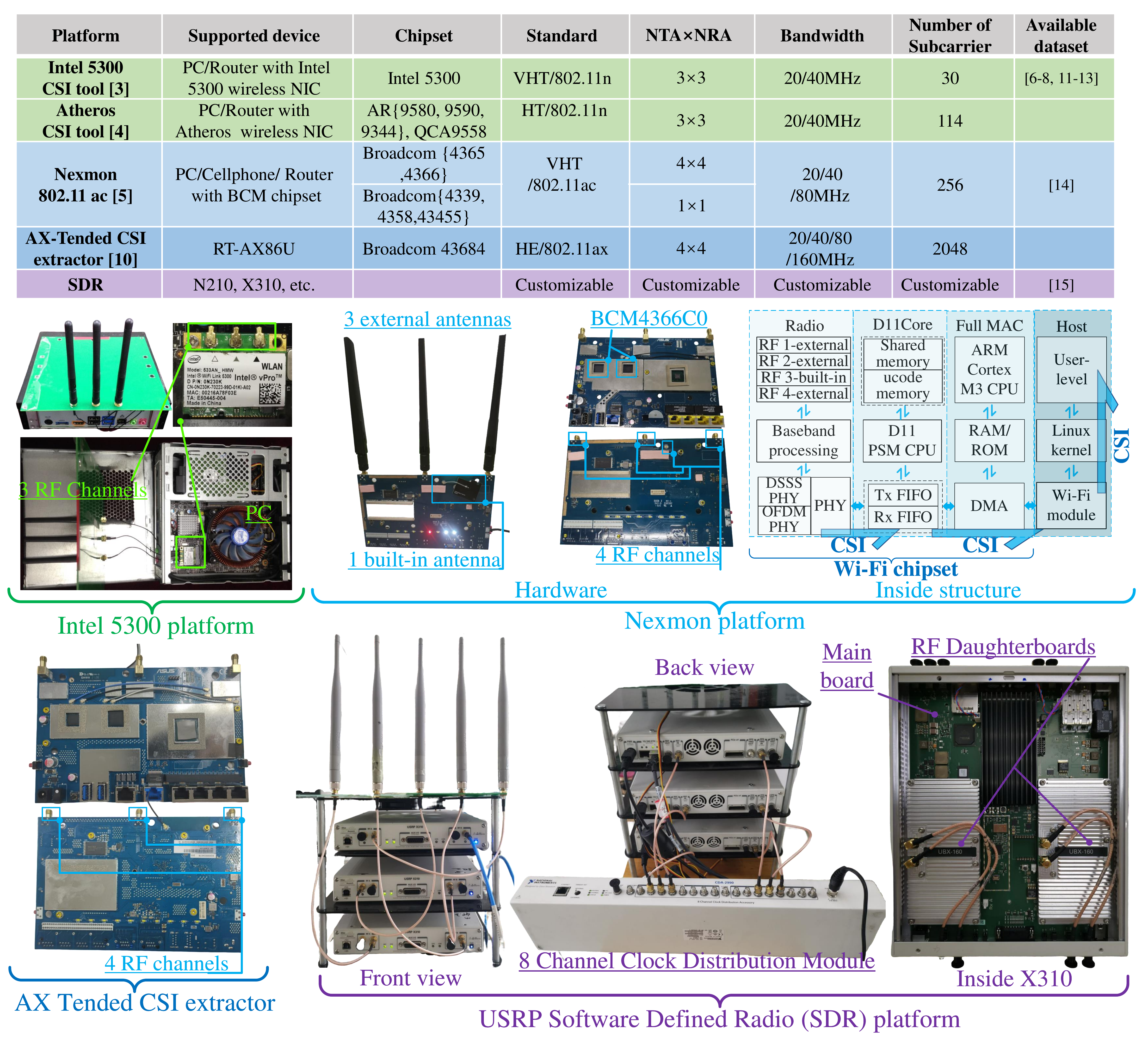}%
	\caption{A comparison of existing platforms. Here, NTA means the number of transmitting antenna, NRA denotes the number of receiving antenna. }
	\label{HDW5}
\end{figure*}


\subsection{The available datasets}
Unlike the natural language processing and computer vision communities that have large public datasets, there are few public datasets in the wireless sensing community. The reason for this is that radio waves are mainly used for communication in various devices. However, with the advances of wireless research in recent years, the important role of communication signals, such as WiFi, in wireless sensing has gradually been recognized, and a number of datasets are now publicly available. The authors in~\cite{ma2018signfi} propose SignFi to recognize language gestures using WiFi. In this work, CSI traces are collected to evaluate SignFi in the lab and home environments. Specifically, the dataset includes 8,280 gesture instances, 5,520 from the lab and 2,760 from the home, for 276 sign gestures. Another dataset for gesture recognition is called WiDar 3.0~\cite{zheng2019zero}, which contains CSI measurements collected from three indoor scenarios, i.e., an empty classroom, a spacious hall, and an office room. Prior to this work, the authors also present WiDar 1.0/2.0~\cite{qian2017widar,qian2018widar2} for passive localization and tracking. 

Moreover, as an important requirement in Metaverse services, activity recognition has also been widely studied. In~\cite{yousefi2017survey}, the authors employ Intel 5300 based equipment to collect CSI data corresponding to 6 activities in an indoor environment and carried out experiments to analyze the activity recognition via different ML models. With the same CSI tool, a dataset named WiAR is presented, which contains CSI data of sixteen activities operated by ten volunteers in three indoor environments \cite{guo2019wiar}. In additoin to the above datasets, authros in \cite{schafer2021human} provide a dataset collected by using Raspberry Pi 3B+, Pi 4B, and Asus RT-AC86U routers with the Nexmon tool. Using the data gathered, the author performs human activity recognition experiments via a deep neural network. They discover that the results are better than those of earlier studies, indicating that the increase in bandwidth could boost the overall sensing performance. Unfortunately, as far as we know, this is the only open-source dataset based on Nexmon so far, and it can only be used for human activity recognition study. Aside from the datasets listed above, a dataset obtained via the SDR platform with a wheeled robot is presented in~\cite{gassner2021opencsi}. This one is mainly used for indoor fingerprinting localization research.

In Fig.~\ref{HDW5}, we summarize the platforms and available datasets, and present pictures of some platforms' hardware. From the above discussion and Fig.~\ref{HDW5}, we can see that most available datasets are collected with Intel 5300 CSI tool. Such datasets are limited in terms of bandwidth, quantization accuracy, etc., thus hindering researchers and practitioners from conducting more in-depth wireless sensing investigations to support Metaverse services. To facilitate the design of more advanced ML as well as signal processing algorithms for human sensing in the physical world, it is imperative to creat an open-source dataset based on the latest platforms.
\section{Data collection, observation, and processing}
This section provides a detailed introduction to the data collection scenario, hardware, and the collection method.
On this basis, the phase shift problem observed in the collected data is thoroughly analyzed. Such phase shift would directly deteriorate the sensing performance of human in the physical world, thus adversely affecting the avatar construction. Therefore, we further propose a novel scheme to detect and correct this shift.
\subsection{Hardware description}
Data is collected via the Asus AP equipped with the Nexmon tool and Broadcom card, which has the structure shown in Fig~\ref{HDW5}. The full structure contains two main parts, i.e., the Host and the chipsets with the FullMAC architecture. In the Host, the Linux Kernel sets up the radio and performs data trading through the direct memory access (DMA) interface. In the chipset, all 802.11-specific functions are managed. One can see that chipset includes three parts, which are the radio layer, D11 Core, and Full MAC.

The bottom layer is the wireless front-end physical interface, which mainly focuses on the tasks in the RF stage, such as detecting data packets through correlation and energy detection, amplifying radio-frequency signals, up and down conversion, etc. These operations are completed in the baseband. The second layer is the microcontroller D11 Core, which executes ucode and also manages time-sensitive tasks such as channel accessing, frame generation, and acknowledging and scheduling via a programmable state machine (PSM), a loop that never ends. The D11 core can communicate with the upper Full MAC through the first-in-first-out (FIFO) queues and read/write operations in the shared memory, while also access to the lower radio layer in a direct manner. The third one is the Full MAC layer, which focuses on the controlling processes connected to frames, like sanity checks and header replacements. It interacts with the upper layers through various interfaces, during which the DMA is used to enhance data transfer; It can directly reach the radio layer and D111 Core. 

The CSI extraction process is completely managed by the D11 Core. Concretely, when the radio layer detects the target frame, the D11 Core shuts down the receiving circuit and freezes CSI for collection. Then the collected CSI data is divided into multiple parts, integrated into the receive header, and pushed to the Full MAC. There, each of these CSI parts is received separately and packaged with other parameters, such as IP and UDP headers, to form a datagram. Finally, Full MAC reports the datagram data to the host, which is then passed to the user after being processed by the Wi-Fi module.
\subsection{Data collection}
\begin{figure}[t]
\centering
\includegraphics[height=7cm]{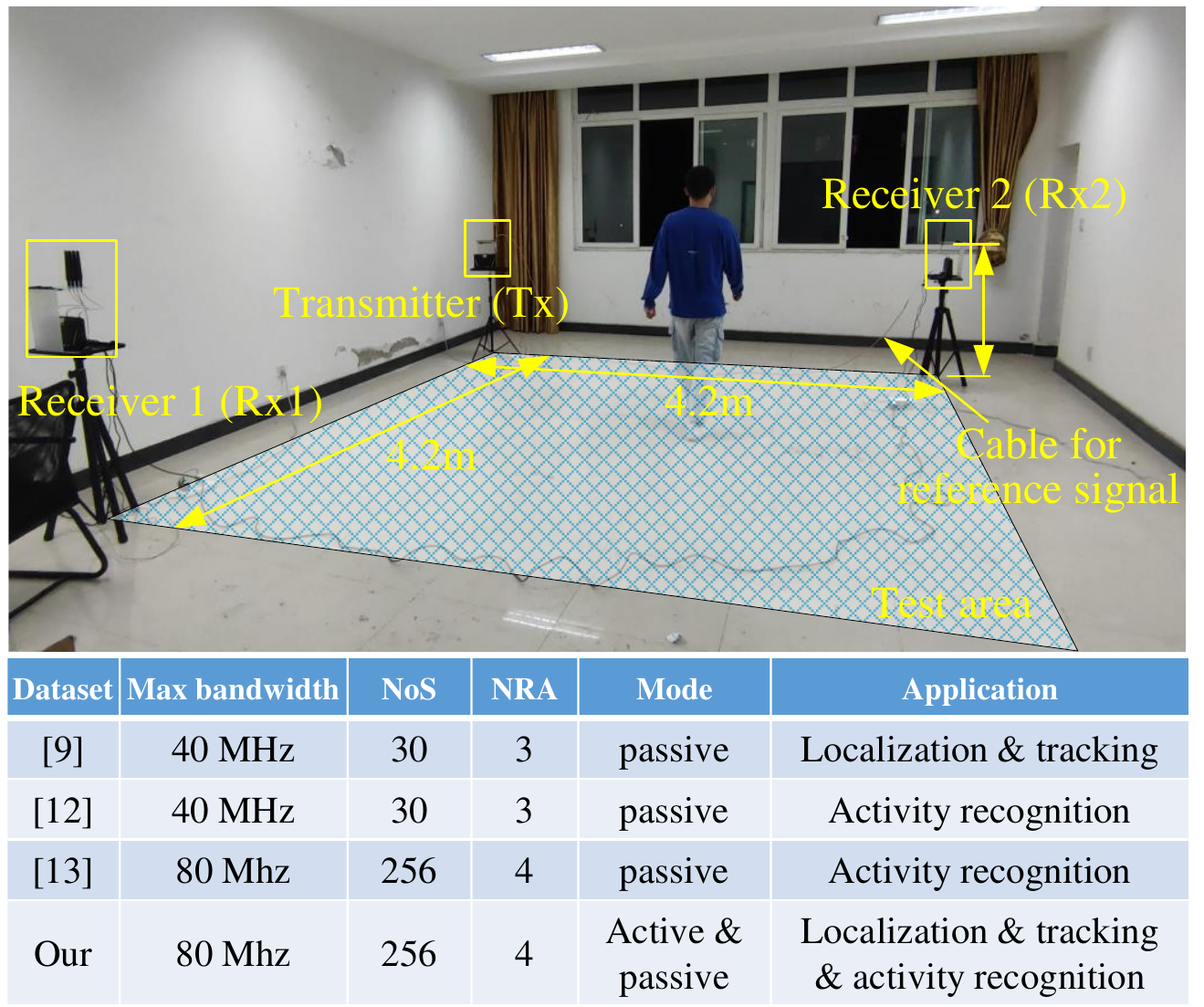} 
\caption{The real-world scenario for data collection and the dataset comparison, where ``NoS'' means the number of subcarries.} 
\label{case} 
\end{figure}

In the physical world, various wireless sensing data shall be gathered in both active and passive manners to support wireless sensing research. By doing so, more support for the Metaverse service provider (MSP), including but not limited to data mining, modeling, rendering, and refreshing, shall be ensured. To this end, we establish two transmission links under the scenario depicted in Fig.~\ref{case} to complete the sensing data collection. During the collection process, the phase difference among the channels at the receiver is first corrected via the power splitter. Then, the transmitter (Tx) and receiver (Rx) are placed as shown in Fig.~\ref{case} to send and receive wireless signals, respectively. Specifically, the data packet rate is 400 Hz, which is extracted by the Ping, the center frequency of the wireless signal is 5.8 GHz, and the bandwidth is 80 MHz, including 256 subcarriers. We compared our dataset with existing data sets from several main aspects such as signal bandwidth.

In this paper, unlike other tests, an antenna of the Rx2 is connected to the transmitter via a cable to receive the signal, which is regarded as the reference signal, while the other three antennas are used to obtain multipath signals from free space. Given that these RF channels share the same processing procedures, the phase offset introduced by synchronization is the same for the CSI obtained from four channels. As a result, using our dataset, the phase error in the CSI corresponding to the multipath signals can be eliminated entirely by the conjugate multiplication operation. This configuration enables more precise analysis of multipath signals from the free space, such as the absolute ToF and Doppler estimation.

\subsection{Data analysis and processing}
Leveraging the device and method presented above, abundant of sensing data is collected. Yet it is observed from the obtained data that the CSI phase difference between the antennas is unsteady. Such an issue would adversely affect sensing performance, particularly the sensing taskes that involve CSI phase, such as human tracking, and finally misleading the construction of avatars in the Metaverse.

\begin{figure}[t]
\centering
\includegraphics[height=8.5cm]{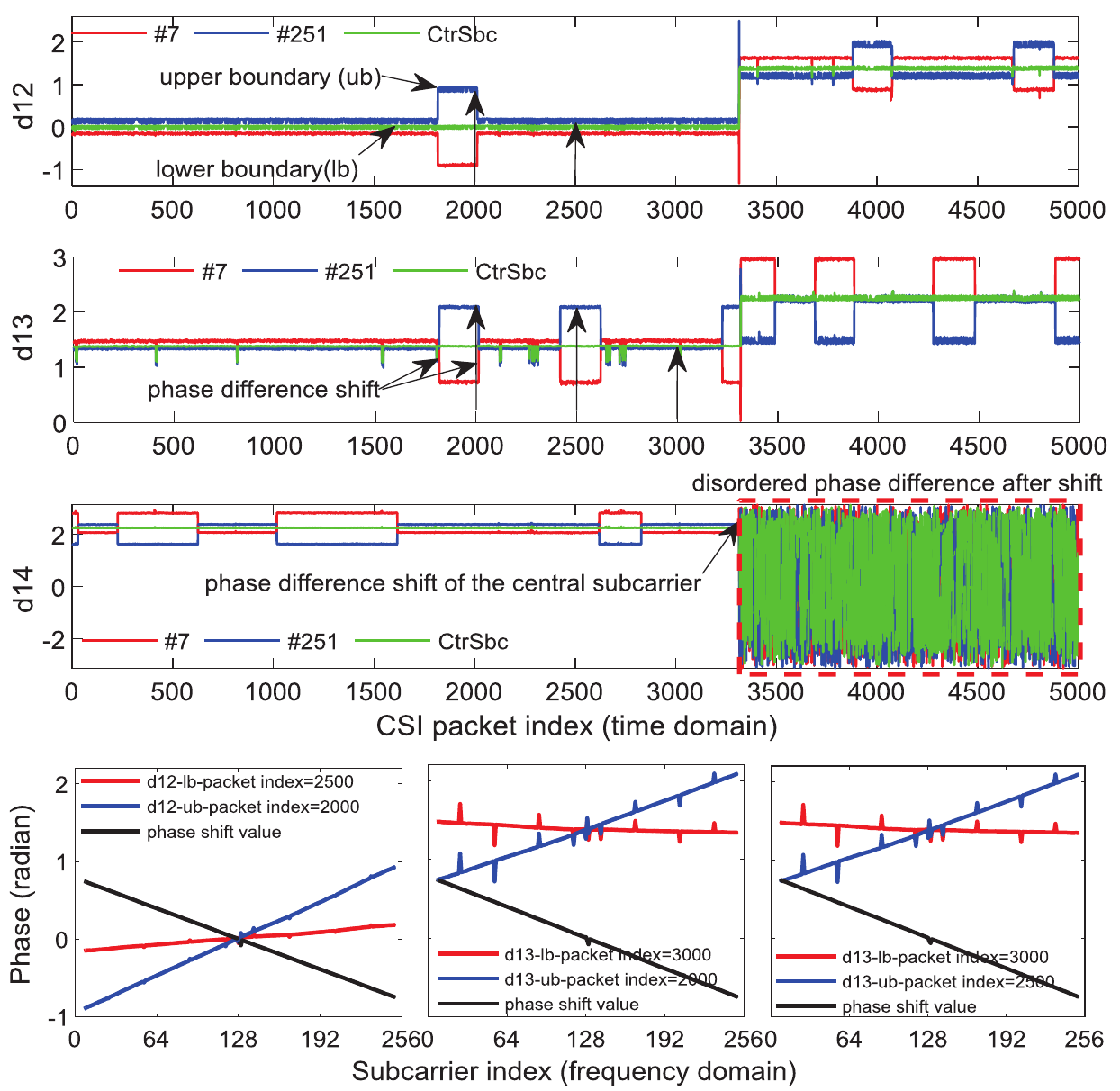} 
\caption{The shift of CSI phase difference between antennas.} 
\label{phase sihft} 
\end{figure}

In particular, we find that the CSI phase difference between the receiving antennas would shift randomly in the time domain, as shown in Fig.~\ref{phase sihft}, where \#7 denotes subcarrier 7, CtrSbc means the central subcarrier, and d12 represents the phase difference between antennas 1 and 2. According to our observations, such shift can be divided into two cases. In the first case, the phase difference between the receiving antennas corresponding to the central subcarrier does not change, while the rest of the subcarriers undergo varying degrees of shift. In the second case, the phase differences between the receiving antennas of all subcarriers are shifted. For the specific shift features, a more in-depth discussion is as follows.
\begin{itemize}
\item In the first case, for a given subcarrier, the phase shift value is a constant. That is, the difference between the upper and lower boundaries never change with time. Meanwhile, in the frequency domain, the shift is linearly related to the frequency of the subcarrier, as the fourth subfigure shows. 

\item The phase shift is the same for symmetrical subcarriers around the center frequency, such as subcarriers 7 and 251.

\item The phase difference between some antennas would become chaotic and hard to fix, as shown in Fig.~\ref{phase sihft}, after the phase difference corresponding to the central subcarrier is shifted. Fortunately, this does not happen very often.
\end{itemize}

In response to the above issues, this paper proposes a straightforward and effective detection and repair scheme, which contains two steps. First, a sliding window is constructed to detect whether the center subcarrier is shifted. Specifically, we regard the phase at the center of the sliding window as the point to be detected, and calculate the difference between the phase at central position and the rest phases in the sliding window. Then the phase shift is believed to happen when the difference exceeds the threshold. Once the shift detected, it is necessary to analyze whether the data can be used according to the situation. Secondly, if the central subcarrier does not shift, the sliding window is utilized to detect whether the other subcarriers undergo shift. Since such shift of subcarrier with larger frequency is more distinct, this detection is best done on subcarrier with larger index. Once identified, proper adjustment is carried out. If the lower boundary is used as the standard, the adjustment can be achieved by subtracting the shift value from the phase difference between the antennas. Otherwise, if the upper boundary is the standard, the adjustment is achieved by addition. Through the above steps, the phase shift can be effectively detected and eliminated, allowing the collected data to be used for more in-depth studies which relies on the phase or phase difference.


\section{Case study}\label{S4A}
Based on the original wireless sensing data extracted from the physical world, a vivid avatar can be constructed from the perspective of multiple dimensions and scales in the Metaverse. However, even with the same dataset, there are certain subtle differences in the constructed avatars due to the different models and methodologies used. For example, the avatars respiratory frequency constructed via the CSI amplitude based model and phase based model may be different. Therefore, we analyze the estimation accuracy of the two most important signal parameters for avatar construction, i.e., the AoA and ToF, rather than directly comparing the construction performance of avatar before and after phase shift correction.
\subsection{The impact of phase shift}
\begin{figure}[t]
\centering
\includegraphics[height=10cm]{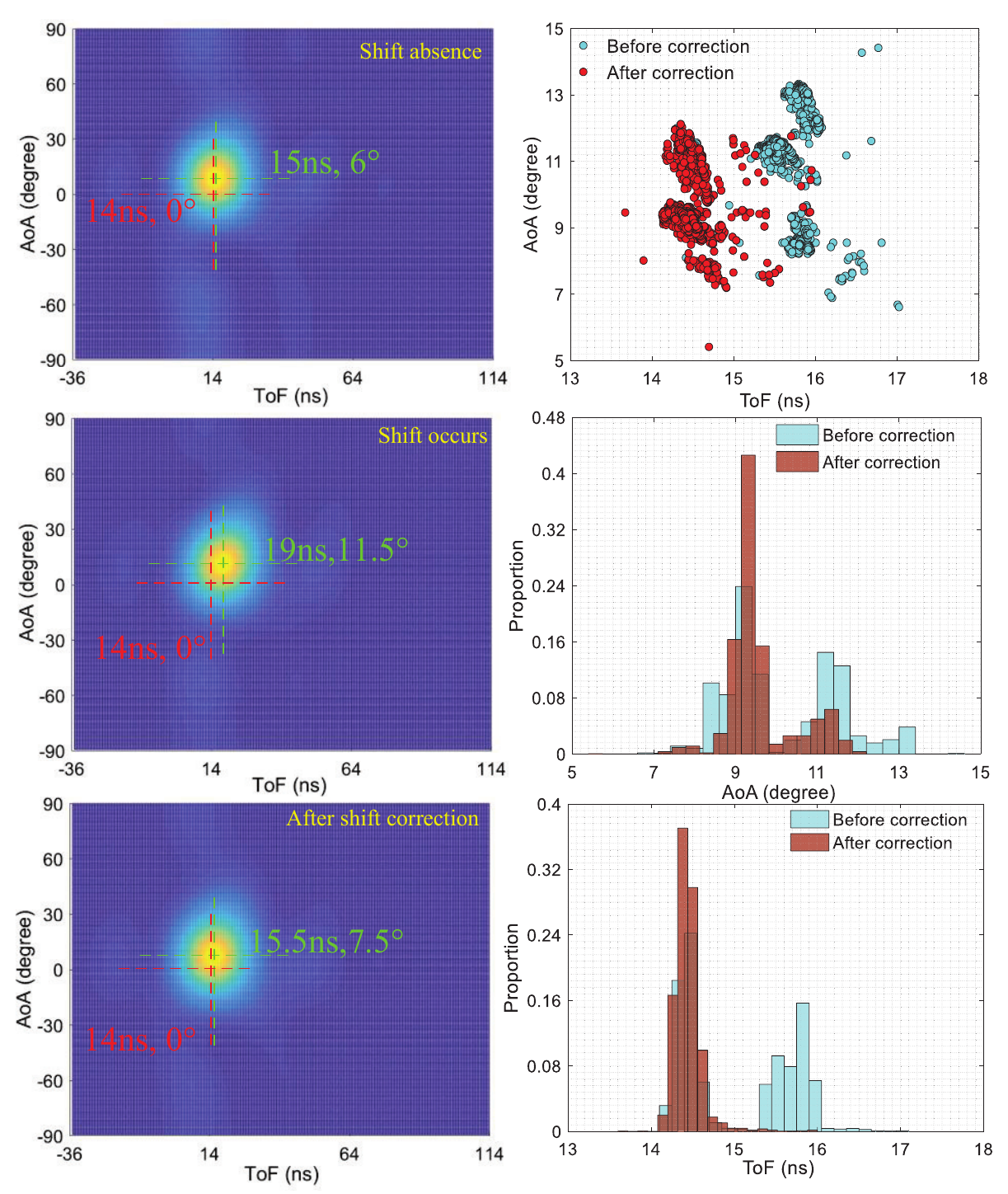} 
\caption{The effect of phase shift on parameter estimation from the perspective of AoA-ToF spectrum and parameter distribution. In the spectrum, the intersection of the red dotted line is the ground truth, and the intersection of the green dotted line is the estimation result.} 
\label{paramet} 
\end{figure}

First, in the active sensing mode, the signal parameter estimation results before and after phase shift correction are compared and analyzed in Fig.~\ref{paramet}. It can be seen from the AoA-ToF spectrum that the peak corresponding to the signal deviates from the right position due to the influence of the phase shift. Taking ToF as an example, in the absence of phase shift, the ToF estimation error is about 1 ns, which is caused by signal noise and the value is reasonable. When the phase shift occurs, such error increases to 5 ns, indicating the shift has a non-negligible impact on the construction of the avatar. Fortunately, after the correction, the spectrum demonstrates that the error is reduced to about 1.5 ns, which is almost the same as the shift-absent case, verifying the effectiveness of the proposed method in phase shift detection and correction. 

On this basis, we accumulate multiple estimation results and compared the parameter distribution before and after correction. It is not difficult to see that the shift drives the parameter distribution deviates from the normal position, resulting in severe negative impact in avatar construction. After correcting the shift, the overall distribution returns to normal, further validating the effectiveness of our method.

\subsection{Signal parameter estimation accuracy analysis}
\begin{figure}[t]
\centering
\includegraphics[height=5.5cm]{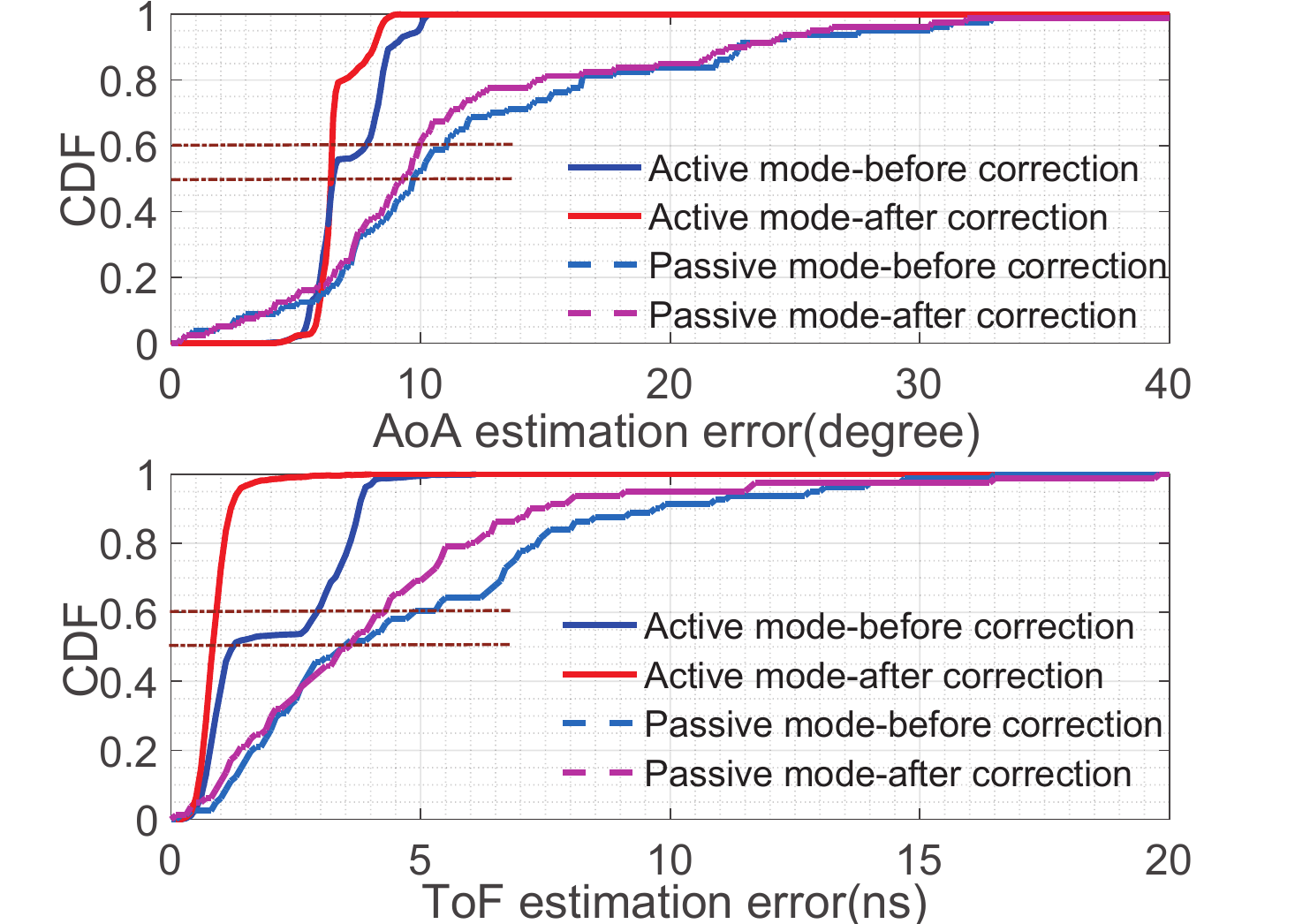} 
\caption{The comparison of signal parameter estimation accuracy before and after phase shift correction.} 
\label{acc} 
\end{figure}

Finally, we statistically analyze the parameter estimation errors introduced by the phase shift in both active and passive modes. As can be seen from Fig.~\ref{acc}, in both active or passive sensing modes, the shift deteriorates the parameter estimation accuracy. Specifically, in the active mode, the shift increases the estimation errors of AoA and ToF with the confidence 0.6 by 1.5 degrees and 2.1 ns, respectively. In passive mode, these two metrics are 1.1 degrees and 1.3 ns, respectively. Relatively speaking, the shift has a greater impact on ToF estimation. Besides that, the curves before and after the shift almost overlap when the confidence level is below 0.5, while there is a significant difference between the curves when confidence level is above 0.5, revealing the estimation error introduced by shift is large.

\section{Future Directions}
\subsection{Wireless sensing data enhancement}
Metaverse avatar construction involves various wireless sensing tasks. However, due to the limitations of the experimental platform, the current datasets can support only limited types of sensing algorithm design. The dataset presented in this paper is collected with the Nexmon platform, which is better than the Intel 5300 based platform in terms of signal bandwidth, number of antennas, etc., yet still slightly behind Software Defined Radio (SDR) and AX-CSI. AX-CSI supports the 802.11ax protocol, which enables us to obtain data with a bandwidth up to 160 MHz from $4 \times 4$ MIMO communications. Even better, the SDR platform supports user-specified wireless parameter configuration, which is beneficial for enhancing sensing performance, such as sensing granularity and coverage. Therefore, more public datasets are needed using these platforms, so as to improve the efficiency of avatar construction, and even the building of the environment around the avatar in Metaverse.

\subsection{Wearable device data and processing}
Since Metaverse is a human centric virtual world and its services require a deeper level of user involvement, it is necessary to promote the collection and sharing of wearable device data in addition to wireless sensing data. Here, the wearable sensors include accelerometers, gyroscopes, magnetometers, and so forth. The reason is that these sensors can record in detail some subtle parameters of users in the physical world, which are otherwise difficult to obtain via wireless signal or image, such as pulse, blood pressure, blood sugar, etc. Meanwhile, considering the unique characteristics of wearable device data, it is also essential to develop the corresponding processing and training models, such as weak signal feature extraction, signal trend prediction, etc. Through the above combination, the Metaverse avatar can be built and analyzed at a deeper level, and many more impressive services can be supported, such as assessing and predicting the avatar's attention and psychological and emotional changes, so as to adjust service strategies and improve the QoS.

\subsection{Multimodal data processing}
In the actual construction process of the Metaverse avatar, a large amount of cross-modal data is collected through various methods. Hence, to achieve an efficient fusion, such technology should comprehensively consider the characteristics of different kinds of data. Not only that, the fusion process should not be static, but needs to fully analyze and consider the Metaverse user's service needs, preferences, and situations, and conduct real-time adjustment according to the resource of the service equipment, including the computing and transmission ability of the MSP, and the rendering and refreshing capability of AR devices. This is not only crucial to the Metaverse avatar, but also important for the construction of the surrounding environment. Therefore,the fusion technique is one of the key concerns in the future work.

\section{Conclusions}
In this article, the application of wireless sensing technology in the construction of Metaverse avatars was discussed, and a wireless sensing dataset for avatar construction was presented. First, the existing open-source wireless sensing platforms and datasets were examined and the corresponding pros and cons were analyzed from various perspectives. On this basis, our dataset, collected by the Nexmon platform, was presented, which can be used for wireless human sensing and then assist avatar construction. We not only detailed the platform hardware structure, test scenario, and method of data collection, but more importantly, we analyzed the dataset, focusing on the phase difference shift between the antennas. Aiming at this shift, we proposed a correction method and verified it in the case study with the presented data set. Finally, several directions of future research were discussed, where wireless sensing is an indispensable part of the Metaverse construction.

\bibliographystyle{IEEEtran}
\bibliography{Ref.bib} 

\begin{thebibliography}{10}
\providecommand{\url}[1]{#1}
\csname url@samestyle\endcsname
\providecommand{\newblock}{\relax}
\providecommand{\bibinfo}[2]{#2}
\providecommand{\BIBentrySTDinterwordspacing}{\spaceskip=0pt\relax}
\providecommand{\BIBentryALTinterwordstretchfactor}{4}
\providecommand{\BIBentryALTinterwordspacing}{\spaceskip=\fontdimen2\font plus
\BIBentryALTinterwordstretchfactor\fontdimen3\font minus
  \fontdimen4\font\relax}
\providecommand{\BIBforeignlanguage}[2]{{%
\expandafter\ifx\csname l@#1\endcsname\relax
\typeout{** WARNING: IEEEtran.bst: No hyphenation pattern has been}%
\typeout{** loaded for the language `#1'. Using the pattern for}%
\typeout{** the default language instead.}%
\else
\language=\csname l@#1\endcsname
\fi
#2}}
\providecommand{\BIBdecl}{\relax}
\BIBdecl

\bibitem{wang2022survey}
Y.~Wang, Z.~Su, N.~Zhang, R.~Xing, D.~Liu, T.~H. Luan, and X.~Shen, ``A survey
  on metaverse: {F}undamentals, security, and privacy,'' \emph{IEEE Commun.
  Surv. Tutor.}, to appear, 2022.

\bibitem{du2022exploring}
H.~Du, J.~Wang, D.~Niyato, J.~Kang, Z.~Xiong, D.~I. Kim \emph{et~al.},
  ``Exploring attention-aware network resource allocation for customized
  metaverse services,'' \emph{arXiv preprint arXiv:2208.00369}, 2022.

\bibitem{halperin2011tool}
D.~Halperin, W.~Hu, A.~Sheth, and D.~Wetherall, ``Tool release: {G}athering
  802.11 n traces with channel state information,'' \emph{ACM SIGCOMM Comput.
  Commun. Rev.}, vol.~41, no.~1, pp. 53--53, Jan. 2011.

\bibitem{xie2015precise}
Y.~Xie, Z.~Li, and M.~Li, ``Precise power delay profiling with commodity
  {WiFi},'' in \emph{Proc. 21st Annual Int. Conf. Mobile Comput. Netw.}, 2015,
  pp. 53--64.

\bibitem{gringoli2019free}
F.~Gringoli, M.~Schulz, J.~Link, and M.~Hollick, ``Free your {CSI}: {A} channel
  state information extraction platform for modern {Wi-Fi} chipsets,'' in
  \emph{Proc. Int. Workshop Wireless Netw. Testbeds Expe. Eva. \& Charact.},
  2019, pp. 21--28.

\bibitem{ma2018signfi}
Y.~Ma, G.~Zhou, S.~Wang, H.~Zhao, and W.~Jung, ``Sign{F}i: {S}ign language
  recognition using {WiFi},'' \emph{Proc. ACM Interact. Mobile Wearable
  Ubiquitous Tech.}, vol.~2, no.~1, pp. 1--21, Jan. 2018.

\bibitem{zheng2019zero}
Y.~Zheng, Y.~Zhang, K.~Qian, G.~Zhang, Y.~Liu, C.~Wu, and Z.~Yang,
  ``Zero-effort cross-domain gesture recognition with {Wi-Fi},'' in \emph{Proc.
  ACM Mobisys}, 2019, pp. 313--325.

\bibitem{qian2017widar}
K.~Qian, C.~Wu, Z.~Yang, Y.~Liu, and K.~Jamieson, ``Widar: {D}ecimeter-level
  passive tracking via velocity monitoring with commodity {Wi-Fi},'' in
  \emph{Proc. ACM MobiHoc}, 2017, pp. 1--10.

\bibitem{xx2c-zg31-22}
\BIBentryALTinterwordspacing
X.~Yang, ``Wireless sensing data for metaverse avatar construction,'' 2022.
  [Online]. Available: \url{https://dx.doi.org/10.21227/xx2c-zg31}
\BIBentrySTDinterwordspacing

\bibitem{gringoli2022ax}
F.~Gringoli, M.~Cominelli, A.~Blanco, and J.~Widmer, ``Ax-csi: Enabling csi
  extraction on commercial 802.11 ax wi-fi platforms,'' in \emph{Proceedings of
  the 15th ACM Workshop on Wireless Network Testbeds, Experimental evaluation
  \& CHaracterization}, 2022, pp. 46--53.

\bibitem{qian2018widar2}
K.~Qian, C.~Wu, Y.~Zhang, G.~Zhang, Z.~Yang, and Y.~Liu, ``Widar2.0: {P}assive
  human tracking with a single {Wi-Fi} link,'' in \emph{Proc. ACM MobiSys},
  2018, pp. 350--361.

\bibitem{yousefi2017survey}
S.~Yousefi, H.~Narui, S.~Dayal, S.~Ermon, and S.~Valaee, ``A survey on behavior
  recognition using wifi channel state information,'' \emph{IEEE Communications
  Magazine}, vol.~55, no.~10, pp. 98--104, 2017.

\bibitem{guo2019wiar}
L.~Guo, L.~Wang, C.~Lin, J.~Liu, B.~Lu, J.~Fang, Z.~Liu, Z.~Shan, J.~Yang, and
  S.~Guo, ``Wiar: {A} public dataset for wifi-based activity recognition,''
  \emph{IEEE Access}, vol.~7, pp. 154\,935--154\,945, 2019.

\bibitem{schafer2021human}
J.~Sch{\"a}fer, B.~R. Barrsiwal, M.~Kokhkharova, H.~Adil, and J.~Liebehenschel,
  ``Human activity recognition using {CSI} information with {N}exmon,''
  \emph{Appl. Sci.}, vol.~11, no.~19, p. 8860, 2021.

\bibitem{gassner2021opencsi}
A.~Gassner, C.~Musat, A.~Rusu, and A.~Burg, ``Opencsi: An open-source dataset
  for indoor localization using csi-based fingerprinting,'' \emph{arXiv
  e-prints}, pp. arXiv--2104, 2021.

\end{thebibliography}
 \begin{IEEEbiographynophoto}
 {Jiacheng Wang} is a research associate with the School of Computer Science and Engineering, Nanyang Technological University, Singapore. \\
 \\\textbf{Hongyang Du}
 is currently pursing the Ph.D degree at the School of Computer Science and Engineering, Nanyang Technological University, Singapore. \\
 \\\textbf{Xiaolong Yang}
 is a lecturer with School of Communication and Information Engineering, Chongqing University of Posts and Telecommunications, China.(e-mail: yangxiaolong@cqupt.edu.cn) \\
 \\\textbf{Dusit Niyato} is a professor with the School of Computer Science and Engineering, Nanyang Technological University, Singapore. He is a IEEE Fellow. \\
 \\\textbf{Jiawen Kang}
 is a professor with the School of Automation, Guangdong University of Technology, China. \\
 \\\textbf{Shiwen Mao}
 is a professor with with the Department of Electrical and Computer Engineering,
 Auburn University, Auburn, USA. He is a IEEE Fellow.
 \end{IEEEbiographynophoto}

\end{document}